# Updated design of the CMB polarization experiment satellite LiteBIRD

H. Sugai[40], P. A. R. Ade[62], Y. Akiba[41,52], D. Alonso[62], K. Arnold[75], J. Aumont[11], J. Austermann[72], C. Baccigalupi[27,19,25], A. J. Banday[11], R. Banerji[57], R. B. Barreiro[59], S. Basak[16,27], J. Beall[72], S. Beckman[74], M. Bersanelli[30,22], J. Borrill[70,74], F. Boulanger[9], M. L. Brown[66], M. Bucher[5], A. Buzzelli[29], E. Calabrese[62], F. J. Casas[59], A. Challinor[65,63], V. Chan[4], Y. Chinone[74,40], J.-F. Cliche[1], F. Columbro[26,24], A. Cukierman[74], D. Curtis[74], P. Danto[7], P. de Bernardis[26,24], T. de Haan[70], M. De Petris[26,24], C. Dickinson[66], M. Dobbs[1], T. Dotani[37], L. Duband[6], A. Ducout[40], S. Duff[72], A. Duivenvoorden[61], J.-M. Duval[6], K. Ebisawa[37], T. Elleflot[75], H. Enokida[35], H. K. Eriksen[57], J. Errard[5], T. Essinger-Hileman[71], F. Finelli[20], R. Flauger[75], C. Franceschet[30,22], U. Fuskeland[57], K. Ganga[5], J.-R. Gao[56], R. Génova-Santos[58,60], T. Ghigna[67,40], A. Gomez[7], M. L. Gradziel[17], J. Grain[9], F. Grupp[15,14], A. Gruppuso[20], J. E. Gudmundsson[61], N. W. Halverson[76], P. Hargrave[62], T. Hasebe[37], M. Hasegawa[41,52], M. Hattori[53], M. Hazumi[41,37,40,52], S. Henrot-Versille[12], D. Herranz[59], C. Hill[74,70], G. Hilton[72], Y. Hirota[35], E. Hivon[8], R. Hlozek[4], D.-T. Hoang[68,77], J. Hubmayr[72], K. Ichiki[45], T. Iida[40], H. Imada[12], K. Ishimura[54], H. Ishino[48,41], G. C. Jaehnig[76], M. Jones[67], T. Kaga[37], S. Kashima[46], Y. Kataoka[48], N. Katayama[40], T. Kawasaki[42], R. Keskitalo[70], A. Kibayashi[48], T. Kikuchi[32], K. Kimura[49], T. Kisner[70], Y. Kobayashi[38], N. Kogiso[49], A. Kogut[71], K. Kohri[41], E. Komatsu[13], K. Komatsu[48], K. Konishi[36], N. Krachmalnicoff[27,19,25], C. L. Kuo[73,69], N. Kurinsky[73,69], A. Kushino[44], M. Kuwata-Gonokami[34], L. Lamagna[26,24], M. Lattanzi[21], A. T. Lee[74,70], E. Linder[70,74], B. Maffei[9], D. Maino[30,22], M. Maki[41], A. Mangilli[11], E. Martínez-González[59], S. Masi[26,24], R. Mathon[11], T. Matsumura[40], A. Mennella[30,22], M. Migliaccio[29,24], Y. Minami[41], K. Mistuda[37], D. Molinari[28], L. Montier[11], G. Morgante[20], B. Mot[11], Y. Murata[37], J. A. Murphy[17], M. Nagai[46], R. Nagata[41], S. Nakamura[55], T. Namikawa[65], P. Natoli[28,21], S. Nerval[4], T. Nishibori[39], H. Nishino[50], Y. Nomura[51], F. Noviello[62], C. O'Sullivan[17], H. Ochi[55], H. Ogawa[49], H. Ogawa[37], H. Ohsaki[35], I. Ohta[43], N. Okada[37], N. Okada[49], L. Pagano[28,21], A. Paiella[26,24], D.

Paoletti[20], G. Patanchon[5], F. Piacentini[26,24], G. Pisano[62], G. Polenta[18], D. Poletti[27,19,25], T. Prouvé[6], G. Puglisi[73], D. Rambaud[11], C. Raum[74], S. Realini[30,22], M. Remazeilles[66], G. Roudil[11], J. A. Rubiño-Martín[58,60], M. Russell[75], H. Sakurai[34], Y. Sakurai[40], M. Sandri[20], G. Savini[64], D. Scott[3], Y. Sekimoto[37,41,52], B. D. Sherwin[65,63], K. Shinozaki[39], M. Shiraishi[47], P. Shirron[71], G. Signorelli[23], G. Smecher[2], P. Spizzi[7], S. L. Stever[40], R. Stompor[5], S. Sugiyama[51], A. Suzuki[70], J. Suzuki[41], E. Switzer[71], R. Takaku[34], H. Takakura[33], S. Takakura[40], Y. Takeda[37], A. Taylor[67], E. Taylor[74], Y. Terao[35], K. L. Thompson[73,69], B. Thorne[67], M. Tomasi[30,22], H. Tomida[37], N. Trappe[17], M. Tristram[12], M. Tsuji[47], M. Tsujimoto[37], C. Tucker[62], J. Ullom[72], S. Uozumi[48], S. Utsunomiya[40], J. Van Lanen[72], G. Vermeulen[10], P. Vielva[59], F. Villa[20], M. Vissers[72], N. Vittorio[29], F. Voisin[5], I. Walker[62], N. Watanabe[42], I. Wehus[57], J. Weller[15,14], B. Westbrook[74], B. Winter[64], E. Wollack[71], R. Yamamoto[32], N. Y. Yamasaki[37], M. Yanagisawa[48], T. Yoshida[37], J. Yumoto[34], M. Zannoni[31,22], A. Zonca[75]

1. McGill Univ, Canada
2. Three-Speed Logic, Inc., Canada
3. Univ of British Columbia, Canada
4. Univ of Toronto, Canada
5. APC, CNRS, Univ Paris Diderot, France
6. CEA, France
7. CNES, France
8. IAP, Sorbonne Univ, France
9. IAS, Univ Paris-Sud, France
10. Institut Néel, France
11. IRAP, Univ de Toulouse, CNRS, CNES, UPS, France
12. LAL, CNRS/IN2P3, Univ Paris-Saclay, Orsay, France
13. MPA, Germany
14. MPE, Germany
15. Univ Sternwarte München, Germany
16. IISER-TVM, India
17. National Univ. of Ireland Maynooth, Ireland
18. ASI, Italy
19. IFPU, Italy
20. INAF – OAS Bologna, Italy
21. INFN Ferrara, Italy
22. INFN Milano, Italy
23. INFN Pisa, Italy
24. INFN Roma, Italy
25. INFN Trieste, Italy
26. Sapienza Univ of Rome, Italy
27. SISSA, Italy
28. Univ di Ferrara, Italy
29. Univ di Roma Tor Vergata, Italy
30. Univ of Milano, Italy
31. Univ of Milano Bicocca, Italy
32. AIST, Japan

# Updated design of the CMB polarization experiment satellite LiteBIRD

33. *Dept of Astronomy, Univ of Tokyo, Japan*
34. *Dept of Physics, Univ of Tokyo, Japan*
35. *GSFS, Univ of Tokyo, Japan*
36. *IPST, Univ of Tokyo, Japan*
37. *ISAS, JAXA, Japan*
38. *ISSP, Univ of Tokyo, Japan*
39. *JAXA, Japan*
40. *Kavli IPMU (WPI), Univ of Tokyo, Japan*
41. *KEK, Japan*
42. *Kitasato Univ, Japan*
43. *Konan Univ, Japan*
44. *Kurume Univ, Japan*
45. *Nagoya Univ, Japan*
46. *NAOJ, Japan*
47. *NITKC, Japan*
48. *Okayama Univ, Japan*
49. *Osaka Prefecture Univ, Japan*
50. *RESCEU, Univ of Tokyo, Japan*
51. *Saitama Univ, Japan*
52. *SOKENDAI, Japan*
53. *Tohoku Univ, Japan*
54. *Waseda Univ, Japan*
55. *Yokohama National Univ, Japan*
56. *SRON, Netherlands*
57. *Univ of Oslo, Norway*
58. *IAC, Spain*
59. *IFCA, CSIC-Univ de Cantabria, Spain*
60. *Univ de La Laguna, Spain*
61. *Stockholm Univ, Sweden*
62. *Cardiff Univ, UK*
63. *KICC, UK*
64. *Univ College London, UK*
65. *Univ of Cambridge, UK*
66. *Univ of Manchester, UK*
67. *Univ of Oxford, UK*
68. *Cornell Univ, USA*
69. *KIPAC, USA*
70. *LBNL, USA*
71. *NASA Goddard, USA*
72. *NIST, USA*
73. *Stanford Univ, USA*
74. *Univ of California, Berkeley, USA*
75. *Univ of California, San Diego, USA*
76. *Univ of Colorado, USA*
77. *USTH, Vietnam*

H. Sugai, et al.

**Abstract** Recent developments of transition-edge sensors (TESs), based on extensive experience in ground-based experiments, have been making the sensor techniques mature enough for their application on future satellite CMB polarization experiments. LiteBIRD is in the most advanced phase among such future satellites, targeting its launch in Japanese Fiscal Year 2027 (2027FY) with JAXA's H3 rocket. It will accommodate more than 4000 TESs in focal planes of reflective low-frequency and refractive medium-and-high-frequency telescopes in order to detect a signature imprinted on the cosmic microwave background (CMB) by the primordial gravitational waves predicted in cosmic inflation. The total wide frequency coverage between 34 GHz and 448 GHz enables us to extract such weak spiral polarization patterns through the precise subtraction of our Galaxy's foreground emission by using spectral differences among CMB and foreground signals. Telescopes are cooled down to 5 Kelvin for suppressing thermal noise and contain polarization modulators with transmissive half-wave plates at individual apertures for separating sky polarization signals from artificial polarization and for mitigating from instrumental *1/f* noise. Passive cooling by using V-grooves supports active cooling with mechanical coolers as well as adiabatic demagnetization refrigerators. Sky observations from the second Sun-Earth Lagrangian point, L2, are planned for three years. An international collaboration between Japan, USA, Canada, and Europe is sharing various roles. In May 2019, the Institute of Space and Astronautical Science (ISAS), JAXA selected LiteBIRD as the strategic large mission No. 2.

## 1 Concept of LiteBIRD

It has been suggested since the 1980s [1,2,3] that inflation occurred in the very early, high energy Universe, to resolve remaining issues of the Big Bang theory, such as uniformity, flatness, and monopole problems. Cosmic inflation predicts primordial gravitational wave production, with quantum fluctuations of spacetime as its origin. The LiteBIRD (Lite (Light) satellite for the studies of B-mode polarization and Inflation from cosmic background Radiation Detection) aims to detect signatures of these primordial gravitational waves in the form of specific curl patterns [4,5] of polarization angle distribution of the Cosmic Microwave Background (CMB). The sizes of expected spiral patterns in the sky are characterized by Hubble lengths at the electron scattering eras of the CMB, since primordial gravitational waves

**Updated design of the CMB polarization experiment satellite LiteBIRD**

that enter the horizon then most effectively produce tensor anisotropies. It is therefore essential to cover the whole sky to investigate these large Hubble lengths at the recombination era as well as the reionization era.

LiteBIRD is focused on this point: targeting [6] both the recombination era with the multipole moment $\ell$ between 11 and 200 and of the reionization era with $\ell$ between 2 and 10, optimizing the angular resolution. The other important concepts of this satellite are a warm launch without the requirements of heavy vessels/tanks, and use of multichroic detectors for the effective exploitation of finite focal plane areas. Advantages of measurements from space are being free from atmospheric effects, providing high sensitivity, stability with less systematic uncertainties [e.g.,7], and no restrictions on observing-band selection. Space measurements also give no pickup from the ground. The Sun-Earth L2 point has been selected, since the Sun, the Earth, and the Moon are all located in almost the same direction, which makes it easier to avoid facing them in terms of optical and thermal aspects. Care should be taken, however, on cosmic ray effects [8] because the satellite is more directly exposed to them. Sky observations are planned for 3 years: the presently guaranteed cooling-chain lifetime is 3.5 years, in which 0.5 year is assigned to the transitional period to the normal observation phase on course to L2. A scanning strategy with a combination of boresight spin angle of 50° around the satellite axis and its precession-like rotation around the anti-Sun direction of 45° is used (Fig.1), since this combination provides not only a fairly uniform sky coverage but also the minimization of instrumental systematic uncertainties in polarization measurements.

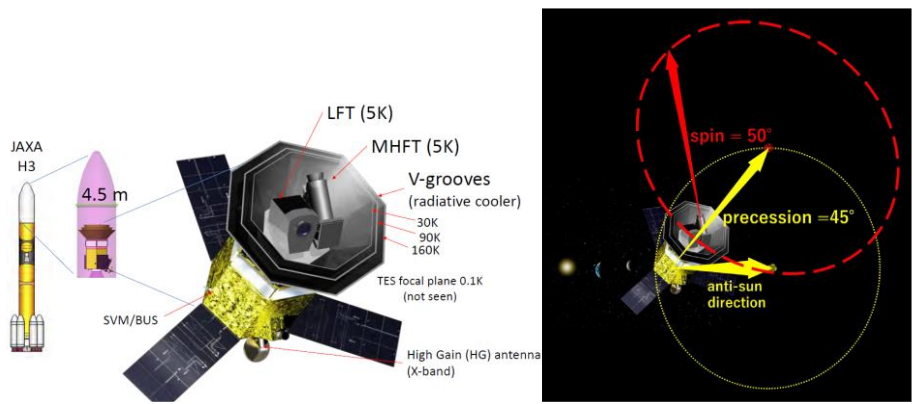

**Fig. 1** *Left:* Conceptual design of LiteBIRD. Warm launch is planned with JAXA's H3 rocket. *Right:* Scans with spin angle 50° and precession angle 45° at L2. (Color figure online)

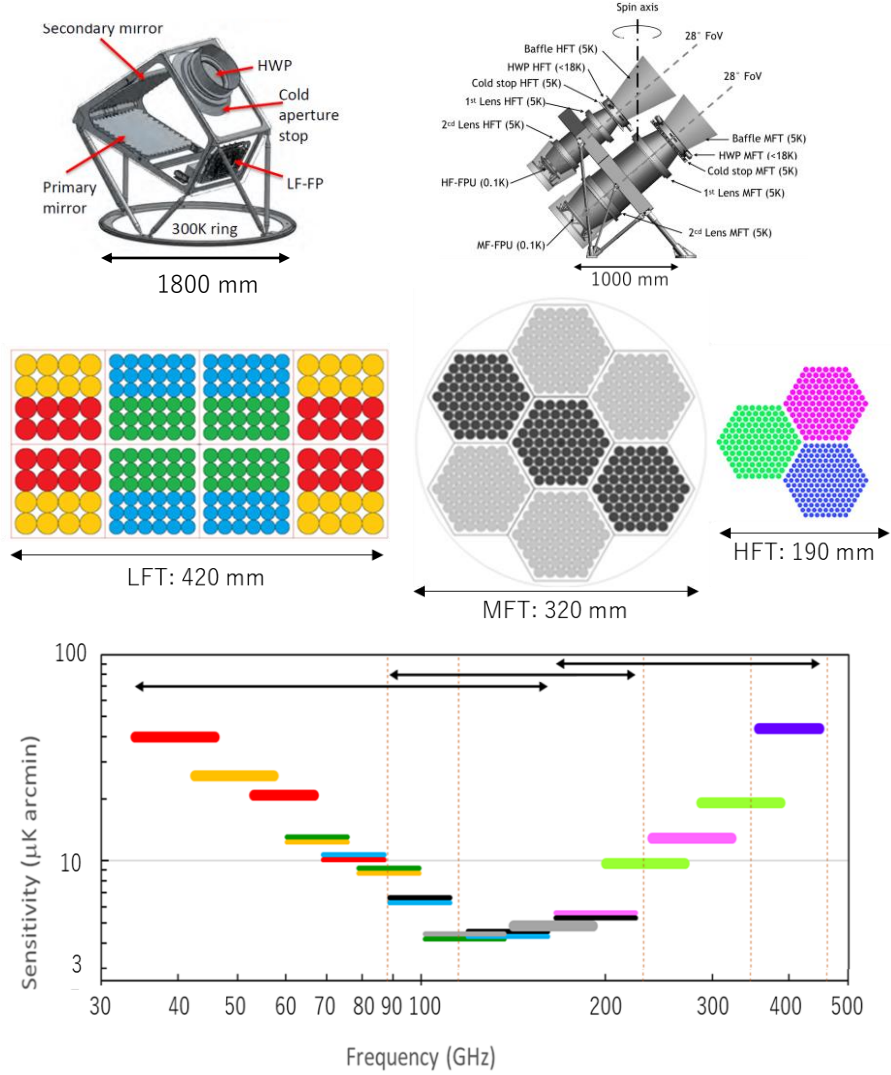

**Fig. 2** *Top:* Designs of LFT (*left*) [9] and MHFT (*right*) [12]. *Middle:* Pixel distribution in LFT (*left*), MFT (*center*), and HFT (*right*) focal planes. Pixel diameters are: 23.6 mm for red (central frequencies of 40, 60, 78 GHz) and orange (50, 68, 89 GHz) pixels; 15.6 mm for green (68, 89, 119 GHz) and light blue (78, 100, 140 GHz) pixels in LFT; 11.6 mm for black (100, 140, 195 GHz) and gray (119, 166 GHz) pixels in MFT; 6.6 mm for pink (195, 280 GHz) and light green (235, 337 GHz) pixels; and 5.7 mm for blue (402 GHz) pixels in HFT. *Bottom:* Predicted LiteBIRD sensitivity. Colors correspond to those shown in the pixel distributions. Frequency ranges for LFT, MFT, and HFT are shown with black arrows. Frequencies of CO $J$=1-0, 2-1, 3-2, 4-3 and HCN $J$=1-0 lines are shown as thin vertical brown dashed lines. Total sensitivity is 2 µK arcmin. (Color figure online)

# Updated design of the CMB polarization experiment satellite LiteBIRD

Table 1. Updated basic parameters and current baseline design for LiteBIRD

| | |
|---|---|
| Mission category | JAXA's strategic large mission |
| Launch vehicle | H3-22L or equivalent |
| Launch schedule | 2027 FY |
| Ground station | JAXA's ground stations (USC, GREAT) |
| Observation period | 3 years |
| Uncertainty of tensor-to-scalar ratio $r$ | $\delta r < 1 \times 10^{-3}$ |
| Multipole moment | $2 \leqq \ell \leqq 200$ |
| Orbit | Second Sun–Earth Lagrangian point L2; Lissajous orbit |
| Scan | Precession angle 45° ($10^{-2}$–$10^{-3}$ rpm); spin angle 50° (0.05–0.1 rpm) |
| Pointing knowledge | < 2.1 arcmin |
| Cooling system | Radiative cooling and mechanical refrigerators (Stirling and JT) without cryogens. Cool in space after launch. ADRs are used to cool the focal plane down to 100 mK |
| Focal-plane detector | Multi-chroic superconducting detector arrays with more than 4000 TES bolometers |
| Sensitivity | 2 μK・arcmin |
| Observing frequencies | 15 bands between 34 and 448 GHz |
| Modulation | Satellite spin and half-wave-plate modulation |
| Data transfer | 9.6 GByte / day |
| Mass | 2.6 ton |
| Electrical power | 3.0 kW |

Table 2. LiteBIRD telescope parameters

| Telescope | Low freq. | Medium freq. | High freq. |
|---|---|---|---|
| Frequency | 34–161 GHz | 89–224 GHz | 166–448 GHz |
| Telescope field of view | 20° × 10° | 28° diameter | 28° diameter |
| Aperture diameter | 400 mm | 300 mm | 200 mm |
| Angular resolution | 70–24 arcmin | 38–28 arcmin | 29–18 arcmin |
| Rotational HWP | 46–83 rpm | 39–70 rpm | 61–110 rpm |
| Number of detectors | 1248 | 2074 | 1354 |

## 2 Status of LiteBIRD

In May 2019, the Institute of Space and Astronautical Science (ISAS), JAXA confirmed that LiteBIRD completed activities planned during Pre-phase A2 (previously called Phase-A1 and dedicated to the concept study and development of key technologies) and selected LiteBIRD as the strategic large mission No. 2, with modifications to the cooling chain and subsequently the focal plane designs: we have chosen to use adiabatic demagnetization refrigerators (ADRs) in series, removing ideas [6,9] of using a JAXA-provided 1-K Joule-Thomson cooler and of having 2-K telescope apertures. To recover the degraded sensitivity with 5-K apertures instead, we have increased the number of detectors. With the international collaboration between Japan, USA, Canada, and Europe, the satellite is planned to be launched with JAXA's H3 rocket in 2027FY (Fig. 1). Tables 1 and 2 summarize the updated basic parameters and current baseline design.

## 3 Current design and technical progresses of LiteBIRD

### 3.1 Cooling chain

With the design modification required, we have selected the following baseline combination for the cooling chain: i) down to 5 K a sunshield and passive cooling with V grooves [10], 15-K pulse tube coolers, and a 4-K J-T with 2ST precoolers; ii) from 5 K to 1.75 K a parallel three-stage ADR for providing continuous cooling at 1.75 K; and iii) from 1.75 K to 100 mK a multi-staged ADR with continuous cooling at 300 mK and 100 mK. Detailed schemes for this cooling chain are described in [11].

3.2 Telescopes and polarization modulator units

The accurate, precise, and robust foreground cleaning requires 15 frequency bands covering a wide range of 34 to 448 GHz. This will be achieved with two kinds of telescopes (Table 2): a reflective one for the lower frequency range (LFT); and refractive ones for medium and high frequency ranges (MHFT [12]). While the LFT uses crossed-Dragone reflective optics [13,14] to minimize effects from multi-reflection among refractive surfaces, the MHFT uses lens optics with the advantage of compactness. For the LFT, the F#3.0 and the crossing angle of 90° have been selected, since this combination suppresses the straylight [9,15] (Fig. 2), and aluminum is used as the mirror material [16]. All the telescopes are cooled down to 5 K for suppressing thermal noise.

Each telescope has a polarization modulator unit (PMU) [17,12,18,19], which consists of a transmissive half-wave plate (HWP) system rotated with a superconductive magnetic bearing (SMB), as the first optical element, close to the aperture stop to distinguish the sky and instrumental polarization components and to reduce the instrumental $1/f$ noise. The SMB enables contactless operation of the HWP system, which minimizes heat generation, through magnetic levitation by using YBCO bulk and permanent SmCo magnets. The LFT HWP system consists of multi-layer stacked sapphire HWPs with optical axes shifted relative to each other to achieve high polarization efficiency over the wide frequency range [20], while the MHFT employs the metal-mesh HWPs [19]. At the top and bottom surfaces of the LFT HWP system, we adopt anti-reflection with sub-wavelength structure produced by laser machining. A cryogenic testbed is designed to study interactions, such as between space-optimized detectors and the PMU, consisting of a magnetically levitated and rotating HWP system [21].

3.3 Detectors and readout

Transition-edge sensor (TES) bolometers are used to form multichroic pixels, based on maturity from successful ground-based and balloon experiments [22]. TES bolometers are coupled with a silicon lenslet and a sinuous antenna for individual broadband pixels for LFT and MFT, while silicon platelet-based corrugated horn and orthomode transducers are used for HFT [23,24]. The noise equivalent power of TES bolometer is proportional to

**Updated design of the CMB polarization experiment satellite LiteBIRD**

$\sqrt{(P_{sat} T_b)}$, with the $T_c/T_b$ ratio being optimized. Here $P_{sat}$, $T_c$, and $T_b$ are the saturated power, transition temperature, and thermal bath temperature, respectively. The TES parameter optimization for LiteBIRD, towards low saturation-power detectors for the satellite environment, has been carried out, as well as a sensor impedance that readily couples to the frequency domain multiplexer. The electro-thermal time constants are controlled by slowing down with the additional heat capacity. Details of these experimental results are described in [25]. Cosmic-ray mitigation methods have been developed by reducing propagation with palladium structures to absorb phonons and with the removal of bulk silicon to block phonons [26]. Readout will be carried out through digital frequency multiplexing, with 68 bolometers connected to each SQUID array amplifier [27,28]. The updated focal-plane designs and the expected sensitivities are shown in Fig. 2.

**4. Foreground cleaning and systematic uncertainty studies**

The focal-plane designs have been determined through iterations with sensitivity calculations based on thermal studies [10], as well as with foreground cleaning (by using methods described in [29,30,31,32]) and systematic uncertainty studies. The baseline foreground-cleaning results have been obtained based on an eight-dimensional parameterization of the foregrounds, including synchrotron power-law spectral index plus curvature and effective temperature of dust, plus power-law index of dust emissivity for $Q$ and $U$ Stokes parameters, each of which is allowed to vary across the sky. Systematic uncertainties have also been studied [33], including beam systematics [14], instrumental polarization and HWP harmonics, polarization efficiency, relative and absolute gain, pointing, polarization angle, time correlated noise, cosmic ray glitches, bandpass mismatch [34], transfer function, non-linearity, and non-uniformity in HWP, with realistic ground/inflight calibration methods taken into account [e.g.,15,35]. Through all these studies, it has been shown that the updated focal-plane designs satisfy LiteBIRD's full success of the total uncertainty in the tensor-to-scalar ratio of less than 0.001, with its uncertainty budgets distributed comparably into a statistical part (including foreground residual and lensing B-mode), a systematic part, and a margin.

**Acknowledgements** This work was supported by World Premier International Research Center Initiative (WPI), MEXT, Japan, by JSPS Core-to-Core Program, A. Advanced Research Networks, and by JSPS KAKENHI Grant Numbers JP15H05891, JP17H01115, and JP17H01125. The Italian contribution to the LiteBIRD phase A is supported by the Italian Space Agency (ASI Grant No. 2016-24- H.1-2018) and the National Institute for Nuclear Physics (INFN). The French contribution to the LiteBIRD phase A is supported by the Centre National d'Etudes Spatiale (CNES), by the Centre National de la Recherche Scientifique (CNRS) and by the Commissariat à l'Energie Atomique (CEA). A Concurrent Design Facility study focused on the MHFT and Sub-Kelvin coolers has been led by the European Space Agency (ESA). The Canadian contribution to LiteBIRD is supported by the Canadian Space Agency. The US contribution is supported by NASA grant no. 80NSSC18K0132.

**Updated design of the CMB polarization experiment satellite LiteBIRD**